\documentclass[aps,pre,floatfix]{revtex4}
\usepackage{epsf}
\usepackage{subfigure}
\usepackage{amsmath}
\usepackage{amssymb}
\usepackage{graphicx}
\usepackage{epstopdf}
\newcommand{\mean}[1]{\left \langle #1 \right \rangle}

\begin{document}

\newcommand{\be}{\begin{equation}}
\newcommand{\ee}{\end{equation}}
\newcommand{\bea}{\begin{eqnarray}}
\newcommand{\eea}{\end{eqnarray}}
\newcommand{\cum}[1]{\left \langle \left \langle #1 \right \rangle 
\right \rangle}

\title{\bf Quantum work relations and response theory}

\author{David Andrieux and Pierre Gaspard}
\affiliation{Center for Nonlinear Phenomena and Complex Systems,\\
Universit\'e Libre de Bruxelles, Code Postal 231, Campus Plaine,
B-1050 Brussels, Belgium}

\begin{abstract}
A universal quantum work relation is proved for
isolated time-dependent Hamiltonian systems in a magnetic field
as the consequence of microreversibility.
This relation involves a functional of an arbitrary observable.
The quantum Jarzynski equality is recovered in the case this 
observable vanishes.
The Green-Kubo formula and the Casimir-Onsager reciprocity relations
are deduced thereof in the linear response regime.
\end{abstract}


\maketitle

Nonequilibrium work relations have recently attracted much interest 
\cite{J97,C99}.
They provide relations for the work dissipated in time-dependent 
driven systems,
independently of the form of the driving. They are of great interest
to evaluate free energies under general nonequilibrium conditions
and they provide new methods to study nanosystems. In the nanoscopic world,
the extension of these classical relations to quantum systems
is of particular importance and different approaches have been proposed.

A first scheme was introduced by Kurchan \cite{K00}.
In this framework, a measurement of the system state
is performed at the initial time. In the sequel, the system
is perturbed by a time-dependent Hamiltonian before performing
another measurement at the final time. The random work performed
on the system is associated with the energy difference between
the final and initial eigenstates. This setup leads to the
quantum extension of Jarzynski equality and Crooks fluctuation theorem
\cite{T00,M05,TLH07,TH07}. Another possibility is to introduce
a quantum work operator which measures the energy difference \cite{MT03},
in which cases quantum corrections to the fluctuation theorem must be taken
into account.  On the other hand, quantum fluctuation theorems have
been obtained in suitable limits where the dynamics admits a Markovian
description, allowing in particular the applications to nonequilibrium
steady states \cite{M03,DM04,EM06,HEM06,EHM07E,EHM07B}.
Yet, the connection between the quantum work relations and response
theory is still an open question even in the linear regime.

The purpose of the present paper is to derive a new type of work relations
which involves a functional of an arbitrary observable. This generating
functional can be related to another functional but averaged over the 
time-reversed process.
This new work relation turns out to be of great generality since we can
recover known results such as Jarzynski equality as special cases.
Furthermore, this universal work relation allows us to formulate the 
response theory,
to derive the quantum linear response functions, the quantum 
Green-Kubo relations \cite{G52,K57},
as well as the Casimir-Onsager reciprocity relations \cite{O31,C45}
in the regime close to the thermodynamic equilibrium.

{\it Functional symmetry relations.} We suppose that the system is described
by a Hamiltonian operator $H(t;\mathcal{B})$ which depends on the time $t$ and
the magnetic field $\mathcal{B}$. The time-reversal operator $\Theta$ is an antilinear
operator such that $\Theta^2=I$ and which has the effect of changing the
sign of all odd parameters such as magnetic fields:
\bea
\Theta H(t;\mathcal{B}) \Theta = H(t; -\mathcal{B}) \, .
\label{TR}
\eea

We first introduce the {\it forward process}.
The system is initially in thermal equilibrium at the inverse temperature
$\beta = 1/k_{{\rm B}}T$. The initial state of the system is described by
the canonical density matrix
\bea
\rho(0)=\frac{{\rm e}^{ -\beta H(0;\mathcal{B})}}{Z(0)} \, ,
\label{rho(0)}
\eea
where the partition function is given in terms of the corresponding
free energy $F(0)$ by $Z(0)={\rm tr} \, {\rm e}^{ -\beta H(0;\mathcal{B})}={\rm 
e}^{ -\beta F(0)}$.
Starting from this equilibrium situation at the initial time $t=0$,
the system evolves until some final time $t=T$ under the Hamiltonian dynamics.
The corresponding forward time evolution is defined as
\bea
i\hbar \frac{\partial}{\partial t} U_{\rm F} (t;\mathcal{B}) = H(t;\mathcal{B})U_{\rm F} (t;\mathcal{B}) \, ,
\label{U}
\eea
with the initial condition $U_{\rm F} (0;\mathcal{B})=I$ \cite{AGT08}.
In the Heisenberg representation, the observables evolve according to
\bea
A_{\rm F}(t)=U^{\dagger}_{\rm F}(t) \, A\,  U_{\rm F}(t)
\eea
which also concerns the time-dependent Hamiltonian
\bea
H_{\rm F}(t)=U^{\dagger}_{\rm F}(t) H(t;\mathcal{B}) U_{\rm F}(t) \, .
\eea
The average of an observable is thus obtained from
\bea
\mean{A_{\rm F}(t)}= {\rm tr} \, \rho(0) A_{\rm F}(t) \, .
\eea
We note that the dependence on the magnetic field is implicit in 
these expressions.

The {\it backward process} is introduced similarly but in the 
magnetic field reversed.
The system is perturbed according to the time-reversed protocol $H(T-t; -\mathcal{B})$,
starting at the initial time $t=0$ from the density matrix
\bea
\rho(T)=\frac{{\rm e}^{ -\beta H(T;-\mathcal{B})}}{Z(T)} \, ,
\label{rho(T)}
\eea
where the free energy $F(T)$ is given in terms of the partition function
according to $Z(T)={\rm tr} \, {\rm e}^{ -\beta H(T;-\mathcal{B})}={\rm e}^{ 
-\beta F(T)}$.
The system ends at time $t=T$ with the Hamiltonian $H(0; -\mathcal{B})$.
The evolution operator of the backward process is defined as
\bea
i\hbar \frac{\partial}{\partial t} U_{\rm R} (t;\mathcal{B}) = H(T-t;\mathcal{B})U_{\rm 
R} (t;\mathcal{B}) \, ,
\label{UR}
\eea
with the initial condition $U_{\rm R} (0;\mathcal{B})=I$ \cite{AGT08},
and is related to the one of the forward process by the following
\\

{\bf Lemma:}
{\it The forward and backward time evolution operators are related to 
each other according to}
\bea
\Theta U_{\rm F} (T-t;\mathcal{B}) U^{\dagger}_{\rm F} (T;\mathcal{B}) \Theta = U_{\rm R} 
(t;-\mathcal{B}) \, ,
\label{lUUl}
\eea
{\it where $t$ is an arbitrary time $0 \leq t \leq T$.}
\\

This lemma is proved by first substituting $T-t$ for $t$ in Eq. (\ref{U}) to get
\bea
-i\hbar \frac{\partial}{\partial t} U_{\rm F} (T-t;\mathcal{B}) = 
H(T-t;\mathcal{B})U_{\rm F} (T-t;\mathcal{B}) \, .
\eea
Multiplying this equation by $U^{\dagger}_{\rm F} (T;\mathcal{B}) \Theta$
from the right and by $\Theta$ from the left, we find
\bea
i\hbar \frac{\partial}{\partial t} \Theta U_{\rm F} (T-t;\mathcal{B}) 
U^{\dagger}_{\rm F} (T;\mathcal{B})
\Theta = H(T-t;-\mathcal{B}) \Theta U_{\rm F} (T-t;\mathcal{B}) U^{\dagger}_{\rm F} (T;\mathcal{B}) 
\Theta \, ,
\eea
where we used the antilinearity $\Theta i = -i \Theta$ of the 
time-reversal operator
and its further property (\ref{TR}). This shows that the expression
$\Theta U_{\rm F} (T-t;\mathcal{B}) U^{\dagger}_{\rm F} (T;\mathcal{B}) \Theta$ obeys the
same evolution equation (\ref{UR}) as $U_{\rm R} (t;-\mathcal{B})$. Since they
also satisfy the same initial condition,
$\Theta U_{\rm F} (T;\mathcal{B}) U^{\dagger}_{\rm F} (T;\mathcal{B}) \Theta = U_{R} (0;-\mathcal{B}) = I$,
we have proven Eq. (\ref{lUUl}). QED.\\

With this lemma, we can now demonstrate the
\\

{\bf Theorem:}
{\it Let us consider an arbitrary time-independent observable $A$ with a definite parity under
time reversal: $\Theta A \Theta = \epsilon_A  A$, with $\epsilon_A = \pm 1$.
It satisfies the following functional relation:}
\bea
\mean{{\rm e}^{\int_0^T dt \lambda(t) A_{\rm F}(t)} {\rm e}^{- \beta 
H_{\rm F}(T)}
{\rm e}^{ \beta H(0)}  }_{\rm F, \mathcal{B}} = {\rm e}^{-\beta \Delta F}
\mean{ {\rm e}^{\epsilon_A \int_0^T dt \lambda(T-t) A_{\rm R}(t)} 
}_{\rm R, -\mathcal{B}} \, ,
\label{qw}
\eea
{\it where $\lambda(t)$ is an arbitrary function, while the subscripts ${\rm F}$
and ${\rm R}$ stand for the forward or backward protocol, respectively.
$\Delta F=F(T)-F(0)$ is the difference of the free energies of the initial
equilibrium states (\ref{rho(T)}) and (\ref{rho(0)}) of the backward 
and forward processes.}
\\

In order to prove Eq. (\ref{qw}), we first consider the quantity 
$A_{\rm F}(t)$,
which can be written as
  \bea
  A_{\rm F}(t)= U^{\dagger}_{\rm F}(t) \, A\,  U_{\rm F}(t)
= U^{\dagger}_{\rm F}(T) U_{\rm F}(T) \, U^{\dagger}_{\rm F}(t) \, A \,
U_{\rm F}(t) \, U^{\dagger}_{\rm F}(T) U_{\rm F}(T) = \epsilon_A \,
U^{\dagger}_{\rm F}(T)\,  \Theta \, A_{\rm R}(T-t) \, \Theta \, 
U_{\rm F}(T) \, ,
\label{Ar}
  \eea
where we have inserted the identity $U^{\dagger}_{\rm F}(T) U_{\rm F}(T)=I$
to go at the second equality. At the third equality, we inserted $\Theta^2=I$
between the evolution operators and we used $\Theta A \Theta = \epsilon_A A$
along with Eq. (\ref{lUUl}). The connection is thus established with the
backward process. Integrating over time with an arbitrary function $\lambda(t)$
and taking the exponential of both sides, the previous expression becomes
  \bea
\exp \left(\int_0^T dt \, \lambda(t) \, A_{\rm F}(t)\right) = 
U^{\dagger}_{\rm F}(T)
\, \Theta \, \exp\left(\epsilon_A \int_0^T dt \ \lambda(T-t)
\, A_{\rm R}(t)\right) \Theta \, U_{\rm F}(T) \, ,
  \eea
  after the change of integration variables $t\to T-t$ in the right-hand side.

Starting from the left-hand side of Eq. (\ref{qw}), we get
  \bea
&& {\rm tr} \,  \rho(0) \exp\left(\int_0^T dt \, \lambda(t) A_{\rm F}(t)\right)
\exp[- \beta H_{\rm F}(T)] \exp[ \beta H(0)] \nonumber \\
&&  = \frac{1}{Z(0)} \, {\rm tr} \,  \exp \left(\epsilon_A \int_0^T dt \,
  \lambda(T-t) A_{\rm R}(t)\right) \Theta \exp [- \beta H(T;\mathcal{B})] \Theta 
\nonumber  \\
& & = \frac{Z(T)}{Z(0)} \, {\rm tr} \, \exp \left(\epsilon_A \int_0^T dt \,
\lambda(T-t) A_{\rm R}(t)\right) \rho(T)
  = {\rm e}^{-\beta \Delta F}  \left\langle \exp \left(\epsilon_A \int_0^T dt \,
  \lambda(T-t) A_{\rm R}(t)\right)\right\rangle_{\rm R, -\mathcal{B}} \, .
\eea
We used the invariance of the trace over cyclic permutations as well as the
exponential of Eq. (\ref{Ar}) at the first equality. In the second equality,
we introduced the equilibrium density matrix (\ref{rho(T)}) which is precisely
the initial condition of the backward process. To obtain the last equality,
we used that the partition functions have been expressed in terms of the
corresponding free energies.
This completes the proof of the theorem. QED.\\

We notice that related results have previously been considered in the 
restricted case 
where there is no change in free energy $\Delta F=0$ 
\cite{BK77,S94}.
The present theorem allows us to recover in 
particular the quantum Jarzynski equality 
as a special case of Eq. 
(\ref{qw}) if $\lambda=0$:
\bea
\mean{{\rm e}^{- \beta H_{\rm F}(T)} {\rm e}^{ \beta H(0)}}_{\rm F, \mathcal{B}} 
= {\rm e}^{-\beta \Delta F} \, .
\eea
The factor inside the bracket can indeed be interpreted in the quantum setting
in terms of the work performed on the system during the forward 
process \cite{K00,T00,TLH07,M03}
in spite of the non-commutativity of the energy operators $H_{\rm 
F}(T)$ and $H(0)$
and thanks to the protocol with von Neumann quantum measurements of the energy
at the initial and final times.  It is only in the classical limit 
that both energies
commute and the classical work can be formed as $W_{\rm 
cl}=\left[H_{\rm F}(T)-H(0)\right]_{\rm cl}$.
In this case, both exponentials in the left-hand side of the relation 
(\ref{qw})
becomes $\exp(-\beta W_{\rm cl})$ which is the classical version of 
this relation.

{\it Response theory.} We can obtain different correlation functions 
by taking functional
derivatives of the relation (\ref{qw}) with respect to the arbitrary 
function $\lambda(t)$.
In this way, we can obtain the expression of linear response theory 
from the generalized
symmetry relation (\ref{qw}). For this purpose, we consider a 
perturbation of the form
\bea
H(t) = H_0 - X(t) B \, ,
\eea
where the perturbation $X(t)$ is such that $X(t)=0$ for $t\leq 0$ and 
$X(t)=0$ for $T\leq t$. The observable $B$ is here arbitrary and should not be confused with the magnetic field $\mathcal{B}$.
In order to obtain the linear response of an observable $A$ with 
respect to the perturbation
$- X(t) B$, we take the functional derivative of Eq. (\ref{qw}) with respect
to $\lambda(T)$, around $\lambda=0$. This yields
\bea
\mean{A_{\rm F}(T) {\rm e}^{- \beta H_{\rm F}(T)} {\rm e}^{ \beta 
H_0}  }_{\rm F, \mathcal{B}}
= \epsilon_A  \mean{ A_{\rm R}(0) }_{\rm R, -\mathcal{B}} = \epsilon_A 
\mean{A}_{{\rm eq}, -\mathcal{B}} \, ,
\label{Alin}
\eea
where we used that $\Delta F=0$ since $X(0)=X(T)=0$. Since the reversed process
also starts at equilibrium, the average in the right-hand side is an 
equilibrium average,
albeit with a reversed magnetic field. Nevertheless, we have that
$\epsilon_A \mean{A}_{{\rm eq},-\mathcal{B}} = \mean{A}_{{\rm eq},\mathcal{B}}$ by using 
time reversal.
We now have to calculate the exponentials of the initial and final 
Hamiltonians.
Since, in the Heisenberg representation, the total time derivative of
the Hamiltonian equals its partial derivative, $d H_{\rm F}/d t = 
(\partial H/\partial t)_{\rm F}$, we can write
  \bea
  \exp[ -\beta H_{\rm F}(T) ] = \exp[ -\beta (H_0 + E) ]
   \eea
  with
  \bea
  E= \int_0^T dt  \left(\frac{\partial H}{\partial t}\right)_{\rm F}  =
  - \int_0^T dt \, \dot{X}(t) \,B_{\rm F}(t) = \int_0^T dt \, X(t)\, 
\dot{B}_{\rm F}(t) \, ,
   \eea
  where the last equality follows from an integration by parts. We now 
use the expression
  \bea
  \exp [\beta (P+Q)] \, \exp(-\beta P) =  1+\int_0^\beta du \, \exp 
[u(P+Q)] \, Q \,  \exp(-uP) \, ,
  \eea
  which can be proved by differentiating with respect to $\beta$.
  To first order in $Q$, we may neglect $Q$ in the last exponential 
function, $\exp[u(P+Q)]$.
  Taking $P=-H_0$ and $Q=-E$ and developing to first order in $X$, we get
  \bea
  {\rm e}^{ -\beta H_{\rm F}(T) } {\rm e}^{\beta H_0}
  =  1- \int_0^T dt\, X(t)\int_0^\beta du \, {\rm e}^{-uH_0} 
\dot{B}(t) {\rm e}^{ u H_0} + O(X^2)
  =  1- \int_0^T dt \, X(t)\int_0^\beta du \, \dot{B}(t+i\hbar u) + O(X^2) \, ,
  \nonumber
  \eea
where $B(t)=\exp(iH_0t/\hbar) B \exp(-iH_0t/\hbar)$ since, at first 
order in the driving force, the time evolution proceeds under
the unperturbed Hamiltonian $H_0$.
Inserting this expansion into Eq. 
(\ref{Alin})
and after some manipulations using the time invariance of correlation function
as well as the KMS-like property $\rho A = A(i\hbar \beta) \rho$ \cite{KMS}, we finally find
  \bea
  \mean{A_{\rm F}(T)}_\mathcal{B} = \mean{A}_{{\rm eq},\mathcal{B}}+ \int_0^T dt \, X(T-t) 
\phi_{AB}(t) + O(X^2) \ ,
\label{greenab}
\eea
with the response function
\bea
\phi_{AB}(t)=\int_0^\beta du \ \langle\dot{B}(-i\hbar 
u)A(t)\rangle_{{\rm eq},\mathcal{B}}  \ .
\label{phiAB}
\eea
Equations (\ref{greenab}) and (\ref{phiAB}) are the well-known expressions of
linear response theory in the canonical ensemble, also known as the 
Green-Kubo formula \cite{G52,K57}.
The Casimir-Onsager reciprocity relations for the conductivities 
\cite{O31,C45} are obtained
by taking $A=J_{\mu}/V$ and $\dot B=J_{\nu}$ in terms of the current 
$J_{\mu}= \sum_n e_n \dot{x}_{n\mu}$
and the volume $V$, in which case the time-reversal symmetry implies
$\phi_{\mu\nu}(t;\mathcal{B})=\phi_{\nu\mu}(t;-\mathcal{B})$ and 
$\sigma_{\mu\nu}(\omega;\mathcal{B})=\sigma_{\nu\mu}(\omega;-\mathcal{B})$
for the tensor of conductivities 
$\sigma_{\mu\nu}(\omega;\mathcal{B})=\int_0^{\infty} dt \, {\rm e}^{i\omega t} 
\phi_{\mu\nu}(t;\mathcal{B})$.
Higher-order terms in the expansion can be obtained as well.

  {\it Conclusions.} In this paper, we have obtained a universal 
quantum work relation
  which involves arbitrary observables at arbitrary times. This result 
relates an average
  over the forward process ponderated by the quantum analogue of the 
work to an average
  over the reversed process. By taking functional derivatives, we can 
obtain relations
  for arbitrary correlation functions, which are the consequence of 
microreversibility.
  In the simplest case, it can be used to recover the well-known 
Jarzynski equality.
  On the other hand, we can also straightforwardly derive from the 
universal relation the
  linear response theory of an arbitrary observable. In this regard, 
this relation
  unifies in a common framework the work relations and the response theory, 
thereby opening
  the possibility to obtain further general relations which are valid 
not only close
  to equilibrium but also in the far-from-equilibrium regime.\\

{\bf Acknowledgments.} D.~Andrieux thanks the F.R.S.-FNRS Belgium for 
financial support.
This research is financially supported by the Belgian Federal Government
(IAP project ``NOSY") and the ``Communaut\'e fran\c caise de Belgique''
(contract ``Actions de Recherche Concert\'ees'' No. 04/09-312).



\begin{thebibliography}{99}
\bibitem{KMS} Indeed, the equilibrium density matrix
$\rho$ is related to the unperturbed evolution $U(t) = \exp(-iH_0t/\hbar)$ according to $Z \rho = U^{\dagger} (i\hbar \beta)$.
\end{thebibliography}
\end{document}